# Interplay of spin-orbit coupling and trigonal crystal field enhances superconductivity in LaAlO$_3$/KTaO$_3$ (111)

Long Cheng[1#], Jia Liu[1#], Tongying Liu[2#], Pan Chen[3#], Mingyue Zhang[1], Jiashi Li[2], Shiyu Zhang[2], Fei Ye[1], Qing Wang[1], Weitao Liu[2*], Jian Kang[1,5*], Jiandi Zhang[3,4*], Xiaofang Zhai[1,5*]

[1] School of Physical Science and Technology, ShanghaiTech University, Shanghai 201210, China

[2] Department of Physics, Fudan University, Shanghai 200438, China

[3] Beijing National Laboratory for Condensed Matter Physics, Institute of Physics, Chinese Academy of Sciences, Beijing, 100190 China

[4] College of Physics, University of Chinese Academy of Sciences, Beijing, 100190, China

[5] State Key Laboratory of Quantum Functional Materials, School of Physical Science and Technology, ShanghaiTech University, Shanghai 201210, China

[#]These authors contributed equally.

[*]Correspondence emails: zhaixf@shanghaitech.edu.cn (X.Z.); jiandiz@iphy.ac.cn (J.Z.); wtliu@fudan.edu.cn (W.L.); kangjian@shanghaitech.edu.cn (J.K.)

# Abstract

In conventional superconductors, bulk physical properties typically degrade as the film thickness approaches the two-dimensional (2D) limit [1-4]. Here in the (111) oriented LaAlO$_3$/KTaO$_3$ (LAO/KTO) heterostructure, we demonstrate experimental evidence that reducing the conducting layer thickness at the interface significantly enhances superconducting transition temperature $T_c$, in direct contrast to conventional wisdom. From the sum frequency generation (SFG) spectroscopy and superconducting upper-critical field measurements, both the trigonal symmetry and spin orbit scattering are enhanced with the increased $T_c$. We attribute the enhanced superconductivity (SC) to the synergic interplay between spin-orbit coupling (SOC) and trigonal crystal field, resulting in an enhanced electron-phonon coupling. Furthermore, we show the existence of unconventional SC: the approaching linear temperature dependence of normal state resistance with increasing $T_c$ and the existence of a quantum critical point (QCP) near the superconducting phase. Our findings provide important insight into the underlying mechanism of the strong orientation-dependent KTO interface SC.

# Introduction

The superconducting transition temperature $T_c$ in 2D superconductors are usually lower than the corresponding bulk system because of strong fluctuations [2,5,6]. Consequently, in thin-film superconductors, the $T_c$ typically decreases as the superconducting thickness is reduced, and the transition exhibits Berezinskii-Kosterlitz-Thouless (BKT) behavior [5,6]. The pursuit of enhanced $T_c$ in superconductors with reduced dimension represents a major scientific challenge, with profound implications for both fundamental physics and the development of next-generation quantum materials – particularly 2D interface superconductors [7]. Recently, KTO has emerged as a new platform for exploring 2D interface SC with a record-high $T_c$ of 2 K [8,9], in comparison to the previous model interface superconductors based on STO. Moreover, because Ta exhibits strong atomic SOC, exotic phenomena such as spontaneous rotational symmetry breaking [10,11] and parity mixing of singlet and triplet [12] have been suggested. Recently, coexistence of SC and ferromagnetism has also been reported [13]. These exceptional properties highlight the critical need to unravel the underlying mechanisms driving the 2D SC in KTO heterostructures.

While KTO-based heterostructures share many similarities with their LAO/STO [14-17] counterparts, they exhibit three fundamental distinctions: (1) a strong orientation dependence of SC [18,19]; (2) the SC only exists in the interface without any sign of SC in bulk KTO via electron doping [20,21]; (3) strong atomic SOC of Ta which is at least one order of magnitude higher than Ti in STO. Among these three features, the orientation-dependent SC may present the most fundamental puzzle, as its resolution could provide key insights into the true origin of SC. The reported $T_c$ in (111), (110) KTO-based heterostructure are 2 K [8,9], 1 K [22] respectively, while SC has only been observed on ionic-liquid gated (111) KTO surface with a $T_c$ of 50 mK [23]. Since SC in KTO heterostructures remains robust regardless of the covering film being amorphous or single-crystalline, the superconducting channel resides predominantly within the KTO side of the interface. Previously, the study of superfluid density in KTO (111) interface has found a superconducting gap larger than the BCS weak coupling limit [24]. An ARPES study of all three orientations indicates enhanced electron-phonon coupling (EPC) strength at the (111) interface [19]. To date, the role of crystal symmetry in the superconducting mechanism of the KTO (111) interface remains elusive, yet such studies could provide critical insights into the origin of the KTO interface SC.

Here we present an in-depth investigation into a series of LAO films, grown at different substrate temperatures on trigonal (111) KTO surfaces, where the temperature serves as a gentle knob to tune the superconducting 2DEG at the interface. We find that $T_c$ nearly linearly increases with decreasing conducting layer thickness. Furthermore, as $T_c$ increases, the spin-orbit scattering rate is drastically increased and the trigonal symmetry of interface is continuously enhanced, revealed by the superconducting critical field measurements and the highly interface-sensitive nonlinear SFG spectroscopy, respectively. We attribute the enhanced SC to the fact of interface 2DEG approaching to the surface of trigonal symmetry, which enhances the SOC related electron-phonon scattering so as the EPC. More strikingly, we find that the unconventional nature of SC, manifested by a nearly linear normal-state resistance from 10 to 300 K and a nearby quantum critical point (QCP), becomes more pronounced as the $T_c$ increases. Our results establish a fundamental framework for understanding exotic SC phenomena at KTO interface.

# Results

**Correlating conducting thickness reduction with $T_c$ enhancement of the (111) LAO/KTO interface** The LAO/KTO heterostructures were fabricated by growing LAO overlayers on the KTO (111) substrates using pulsed laser deposition (PLD). A series of LAO/KTO samples were grown with various substrate temperatures ranging from room temperature to 600 °C. The standard Hall bar geometry was adopted for electrical transport measurements. More details about the sample fabrication and transport measurements are described in the Methods section and Extended Fig. E1-E2.

Figure 1a presents the sheet resistance ($R_s$) as a function of temperature in the low-temperature range (0.3 K-10 K) to illustrates the systematic variation of $T_c$ for our LAO/KTO (111) samples. The samples were labeled in descending order according to $T_c$, which is defined by the temperature at which $\frac{R_s(T_c)}{R_s(10K)}$=1/2, with $R_s(10K)$ representing the normal resistance in the plateau region right above the superconducting transition. Sample #1 exhibits the highest $T_c$ of 2.08 K in this work, reaching the highest record observed at the KTO interface [8,9] to our knowledge. Sample #6 has the lowest $T_c$ of 1.24 K in this work. Furthermore, the 2D superconducting character is confirmed by the significantly enhanced upper critical field along the parallel direction ($H_{c2//}$) than that along the perpendicular direction ($H_{c2\perp}$) (Extended Figs. E3-E5) and the Berezinskii-Kosterlitz-Thouless (BKT) transition with the power law $V \propto I^\alpha$ behavior (Extended Fig. E6). Interestingly, the resistance in the large temperature range from 10 K to 300 K show the non-Fermi liquid behavior of $R_s \propto T^N$ (1<N<2), as shown in Fig. 1b. We found $N$ systematically decreases from 1.53 to 1.15 as $T_c$ increases from 1.24 K to 2.08 K. We will discuss later about the implication of this behavior.

Figure 1c shows the 2D carrier density ($n_{2D}$) and the mobility ($\mu$) measured in the normal state at 12 K using standard Hall measurements. It is observed that the $n_{2D}$ fluctuates with increasing $T_c$. Thus, the $n_{2D}$ alone does not explain the $T_c$ enhancement. On the other hand, the mobility $\mu$ monotonically decreases with increasing $T_c$, which is in line with earlier work [9]. However, it is difficult to correlate the physics of $T_c$ enhancement with the mobility change.

To further explore the key factor contributing to the $T_c$ enhancement, we obtain the conductive layer thickness $d_c$ in the normal state ($T$ = 12 K) by performing magnetoresistance (MR) measurements with magnetic field applied both parallel and perpendicular to the sample plane (see Extended Fig. E7). $d_c$ was estimated by the equation

$$\frac{\text{MR}_{\text{IP}}}{\text{MR}_{\text{OP}}} = \left(\frac{d_c}{l}\right)^2, \qquad (1)$$

which is based on the orbital effect from the electrons' cyclotron motion perpendicular to the applied magnetic field [26, 27]. The MR is defined as $\frac{R_s(H)-R_s(H=0)}{R_s(H=0)}$ with IP/OP denoting the applied magnetic field in or out of the sample plane, and $l$ is the mean free path. As shown in Fig. 1d, the conductive layer thickness $d_c$ obtained from MR measurements in the normal state of 12 K exhibits a decreasing trend with increasing $T_c$. The highest $T_c$ corresponds to the lowest $d_c$ of 3.7 nm, while the lowest $T_c$ corresponds to the largest $d_c$ of 10.5 nm. Thus, the quasi-2DEG at the LAO/KTO interface with higher $T_c$ is more spatially confined along the interface normal. Note that the mean free path $l$ was derived using a single-band non-interacting electron model, which typically yields overestimated values in multiband systems. The extent of this overestimation is addressed in the Method section.

The possibility of interfacial interdiffusion induced conduction layer thickness change is ruled out by the atomic-scale cross-sectional scanning transmission electron microscopy (STEM) measurements and the energy-dispersive X-ray spectroscopy (EDS) imaging study. The bottom part of Fig. 1e shows the large-scale high-angle annular dark field (HAADF) image of sample #1, with the highest $T_c$ among all samples, probed along the [110] zone axis. A sharp interface between the KTO (111) substrate and the amorphous LAO layer is observed. The top part of Fig. 1e shows the EDS elemental maps with single atomic resolution, which reveals stronger La interdiffusion than the K, Ta, Al interdiffusion. Further, the line profiles of La/K EDS spectra across the interface (see the top-left in Fig. 1e) quantify the inter-diffusion depth into KTO side is about 1 nm. Similar interdiffusion depth has also been observed in sample #5 with a very low $T_c$ but a very large conducting layer thickness $d_c$ (Extended Fig. E8). Therefore, the cation interdiffusion does not account for the conducting layer thickness variation between samples of higher $T_c$ and those of lower $T_c$.

**Superconducting layer thickness reduction accompanied with SOC enhancement** To further characterize the superconducting layer thickness, isotherm MRs were measured at different temperatures with perpendicular and parallel magnetic fields, respectively. The upper critical fields $H_{c2\parallel}(T)$ and $H_{c2\perp}(T)$ are defined at the field where the resistance is half of $R_N$, and their temperature dependence is shown in Extended Fig. E3-E5. The phenomenological 2D Ginzburg-Landau (G-L) model is used to quantitatively deduce the superconducting coherence length $\xi_{GL}$ and the superconducting thickness $d_{sc}$ [28]. For the perpendicular field, the G-L equation is the following.

$$\mu_0 H_{c2\perp}(T) = \phi_0(1 - T/T_c)/[2\pi(\xi_{GL})^2], \qquad (2)$$

$\phi_0$ is the quantum magnetic flux. For the parallel field, the 2D G-L model yields the following equation.

$$\mu_0 H_{c2\parallel}(T) = \phi_0\sqrt{12(1 - T/T_c)}/[2\pi\xi_{GL}d_{sc}], \qquad (3)$$

As shown in Fig. 2a and b, the normalized upper critical fields $\frac{H_{c2\parallel}(T)}{H_{c2\parallel}(0K)}$ and $\frac{H_{c2\perp}(T)}{H_{c2\perp}(0K)}$ vs. normalized temperature ($T/T_c$) of all six samples collapse well to a single curve near the phase transition region (0.5 <$T/T_c$< 1). Hence the G-L equations nicely fit this region, from which the $\xi_{GL}$ and $d_{sc}$ of all samples are extracted and shown in Fig. 2c, which both decrease with increasing $T_c$. Sample #1 with the highest $T_c$ of 2.08 K has the thinnest $d_{sc}$ of 3 nm, which is slightly less the estimated $d_c$ of 3.7 nm based on Eq. (1). Furthermore, $\xi_{GL}$ is much larger than $d_{sc}$, which is also a signature of the quasi-2D superconductor.

For a 2D superconductor under a parallel magnetic field, the orbital motion and vortices can be generally neglected. Thus, the Zeeman energy dominates the pair breaking. In the BCS weak coupling limit, the Chandrasekhar-Clogston Pauli paramagnetic limit sets the upper bound for the $H_{c2//}$ [29-32], which is given by $H_P = \frac{\Delta_0}{\sqrt{2}\mu_B}$ with $\Delta_0 \approx 1.76k_B T_c$ being the BCS superconducting gap and $\mu_B$ the Bohr magneton. Since the $H_{c2\parallel}$ exceeds our measurement limit at low temperatures especially for those higher $T_c$ samples, we estimate the $H_{c2\parallel}(T = 0)$ using the G-L equation. Figure 2d shows that the G-L estimated $H_{c2\parallel}$ at $T = 0$ K significantly exceeds the $H_P$ and the ratio $H_{c2//}/H_P$ diverges with $1/d_{sc}$. The violation of the Pauli limit can be explained by strong SOC [29, 30], which induces spin randomization and weakens the paramagnetic effect. Thus, the continuous enhancement of $H_{c2//}/H_P$ suggests that the SOC effect is enhanced as $d_{sc}$ decreases.

To further reveal the SOC effect and its relation with SC, the Klemm–Luther–Beasley (KLB) model for 2D superconductors [33,34] has been adopted to deduce the spin-orbit scattering time $\tau_{so}$, and the modeling details are shown in Methods and Extended Fig. E9. Independently, we obtain the transport (elastic) scattering time $\tau_{tr}$, for which both spin-dependent and spin-independent scattering events are included, from the Drude model of $\tau_{tr} = \frac{m_e^* \mu}{e}$ by extending the measured mobility $\mu$ to zero temperature (see Extended Fig. E2c) and taking $m_e^*$ as $0.3 m_e$ [35]. Figure 2e illustrates that both $\tau_{so}$ and $\tau_{tr}$ dramatically decrease with decreasing $d_{sc}$. More importantly, the positive correlation of $\tau_{so} \propto \tau_{tr}$ indicates the Elliott-Yafet (E-Y) type spin relaxation mechanism [36,37]. In the E-Y model, the electron spin is strongly coupled to the crystal lattice momentum and flips during the spin-orbit scattering. In comparison, the D'yakonov-Perel' (D-P) spin relaxation mechanism due to Rashba SOC holds a reversed relation of $\tau_{so}^{-1} \propto \tau_{tr}$ [38,39], which is not reflected by our data. Previously, both indications of E-Y type and D-P type spin relaxations have been observed in electric-gated EuO/KTO (110) interface [40]. The present findings of the (111) KTO interface show the dominance of scattering induced spin relaxation from crystal lattice over Rashba SOC. Moreover, as $T_c$ increases from sample #6 to #1, we observe the $\tau_{so}^{-1}$ increases much faster than the $\tau_{tr}^{-1}$, with the ratio of $\frac{\tau_{so}^{-1}}{\tau_{tr}^{-1}}$ continuously increasing from ~ 4% in sample #6 to ~ 16% in sample #1 (see Fig. 2f). Additionally, we also used the modified Werthamer-Helfand-Hohenberg (WHH) model [41,42] for thin films to repeat the above analysis. Different from the bulk WHH model, the modified WHH model treats the superconducting layer as a finite thickness slab. The obtained $\tau_{so}$ and its relation with $\tau_{tr}$ are shown in Extended Fig. E10, which confirm the above conclusions from the KLB modeling.

**Trigonal lattice symmetry at the interface revealed by SFG** We further studied symmetry properties of the interfacial layers using the nonlinear optical sum-frequency generation (SFG) technique. The principle and setup of SFG were described in detail elsewhere [43, 25] and in the Methods. In a centrosymmetric system, SFG becomes electric-dipole forbidden due to the mutual exclusion between one-photon and two-photon selection rules [43]. It is thus specific to surfaces and interfaces where the inversion symmetry is broken, such as the (111) KTO surface. Experimentally, the SFG signal was generated by an infrared (IR) and a near-infrared (NIR) femtosecond laser beams overlapping spatially and temporally on the KTO heterostructures (top panel of Fig. 3a). The sample can be rotated about the surface normal, with $\varphi$ denoting the azimuthal angle between the beam incident plane and the [1 -1 0] axis. A typical SFG spectrum from the (111) KTO bare surface is presented in Fig. 3b (bottom panel, filled in grey), exhibiting a prominent resonance peak near 107 meV. Similar to the case of SrTiO$_3$ (STO), the 107 meV-mode of KTO is assigned to the surface projection of the highest zone center longitudinal optical phonon mode (LO$_4$) of bulk KTO. The spectra thus provide information about properties of interfacial KTO lattices. The SFG response at the lower frequency side $\sim$ 100 meV is attributed to the Ta-O local vibrational modes associated to oxygen vacancies [25, 44].

Figure 3b presents SFG spectra from two (111) LAO/KTO samples, sample #1 ($T_c$ = 2.08 K) and sample #5 ($T_c$ = 1.47 K), taken at $\varphi = 0°$ (upper panel) and 60° (middle panel), respectively. It is noted that the SFG intensity from the LAO/KTO interfaces were over 30 times stronger than the bare KTO surface, indicating the further lowering of symmetry upon the formation of conduction channel at KTO subsurface. Though the SFG responses from samples #1 and #5 had similar peak intensities, they were markedly distinct in anisotropy, as illustrated in Fig. 3c. For each sample in Fig. 3b, the SFG spectral profile remained essentially the same at different $\varphi$. We therefore plotted the integrated SFG intensity (between 90 and 120 meV) versus $\varphi$ in Fig. 3c of the three samples. Notably, sample #1 (with a higher $T_c$) differs from both the bare substrate and sample #5 (with a lower $T_c$), exhibiting a distinct six-fold pattern as shown in Fig. 3c (bottom panel).

Given that KTO possesses a simple cubic structure with Pm$\bar{3}$m space group (Fig. 4a, left panel), the bulk-like subsurface contribution is isotropic ($\overleftrightarrow{\chi}^{(2)}_{eff,B}$) [25,43,44]. Meanwhile, at the (111) surface, focusing on the topmost Ta layers, the broken inversion symmetry along the [111] orientation leads to a 3m local symmetry (as depicted in Fig. 4a, right panel), and the surface effective susceptibility takes an anisotropic form $\overleftrightarrow{\chi}^{(2)}_{eff,S} \propto \cos 3(\varphi + \varphi_0)$. Thus, the total nonlinear susceptibility combining the surface and bulk-like subsurface contributions yields an SF signal as:

$$I_{SF} \propto \left|\overleftrightarrow{\chi}^{(2)}_{eff,S} + \overleftrightarrow{\chi}^{(2)}_{eff,B}\right|^2 \propto \left|A \cos 3(\varphi + \varphi_0) e^{i\theta} + B\right|^2. \quad (4)$$

Here, A and B are coefficients representing the amplitudes of anisotropic and isotropic terms, respectively. The term $e^{i\theta}$ symbolizes the phase difference between the surface and bulk components. The surface term alone leads to a six-fold symmetry in the scanning $\varphi$ diagram, and the second term alone is isotropic. When both terms are present and comparable in strength, the interference leads to a three-fold $\varphi$ dependence. In Fig. 3c, we observe such three-fold $\varphi$-scan patterns for the KTO bare substrate and the LAO/KTO sample #5, suggesting appreciable electric-field-induce response from thicker conductive layers from the subsurface of KTO. In contrast, the six-fold pattern of LAO/KTO sample #1 indicates the dominance of KTO surface contribution. Fourier transform and nonlinear fitting are employed to determine the A and B coefficients (see details in the Methods section), and the ratio of A/B reflecting the proportion of surface to bulk contributions are found to be 0.12, 0.49 and 0.63 for the KTO bare substrate, LAO/KTO sample #5 and sample #1, respectively. On the right side of Fig. 3c, we schematically show the A to B ratio as represented by the columns named as '*3m*' and 'Iso,' respectively. Therefore, the conducting layer of higher $T_c$ sample experiences stronger influence from the surface trigonal than the bulk cubic symmetry. Such a symmetry crossover from $Pm\bar{3}m$ to *3m* is necessary for the orbital degeneracy lifting of the $t_{2g}$ manifolds (*xy*, *xz*, *yz*) at the (111) surface. The hierarchical structure of 5d orbitals is depicted in Fig. 4b. For the KTO (111) system in this work, the trigonal crystalline field splits the $t_{2g}$ manifold into $a_{1g}$ and $e'_g$ manifolds. Besides, the strong SOC also splits the $t_{2g}$ levels into a $j = 1/2$ doublet and a $j = 3/2$ quadruplet. With both the trigonal field and the SOC, $t_{2g}$ manifold splits into three energy levels and $e_g$ manifold splits into two energy levels, that is, all the degeneracies are lifted [45], which may serve as mediators of electron pairing.

## Discussions

We discuss possible mechanisms responsible for the $T_c$ enhancement together with the unconventional behavior of SC in samples with reduced 2DEG thickness. Since the 2D carrier density does not increases with the $T_c$, the first possible reason is the increase of 3D carrier density by considering the finite thickness. However, previous studies reveal that KTO bulk doped at the similar level [20,21] to the current work, about 0.004 to 0.01 electron per unit cell, is not superconducting. Then we consider the second possible mechanism for the $T_c$ enhancement: the effect of reduced dimensionality. As $d_{sc}$ decreases from 6.4 nm to 3 nm, the system approaches closer to the 2D limit. Using the simple non-interaction single band model, we compare the effective density of states (DOS) at the Fermi level between 2D and 3D as $\frac{g_{E_f}^{2D}}{g_{E_f}^{3D} \cdot d_c}$, which is found to be less than 1 in all samples and thus still cannot explain the observed $T_c$ enhancement.

Next, we consider the role of the band splitting due to the trigonal surface symmetry. As schematically shown in Fig. 4b, all orbitals of the $t_{2g}$ manifold in the surface split into three narrow bands by the combination of the crystal field ($\Delta_{CF}$) and SOC ($\mathcal{L}$). The energy splitting between the two lowest bands is comparable in magnitude to the smaller of two parameters: $\Delta_{CF}$ and $\mathcal{L}$. In density functional theory calculations, $\Delta_{CF}$ is highly dependent on the thickness of the calculated KTO slab and the reported $\Delta_{CF}$ is about 15 to 20 meV [46,47] in a slab composed of two Ta monolayers in (111) orientation, which is much smaller than the reported $\mathcal{L}$ of 0.4 eV [48]. Therefore, the energy splitting between the two lowest lying bands is estimated to be around 15 meV for the topmost Ta layer, while decreases for the deeper layers. Without considering SOC or Fermi surface topology change, such splitting is expected to weaken SC via suppressing interorbital pairing. While what we observed here is the $T_c$ enhancement in samples with stronger trigonal symmetry, indicating such pairing suppression is not important.

Having excluded the above trivial effects typically responsible for $T_c$ change in conventional superconductors, we now consider the nontrivial effects related to the trigonal symmetry and strong SOC. As $T_c$ increases from sample #6 to #1, the spin-orbit scattering rate $\tau_{so}^{-1}$ is enhanced by more than one order of magnitude while the $\tau_{tr}^{-1}$ only increases by 2-3 times. In the E-Y model, the two scattering rates obey the linear relation of $\tau_{so}^{-1} \propto \frac{\mathcal{L}}{E_g} \tau_{tr}^{-1}$, with the $E_g$ and $\mathcal{L}$ denoting the band gap and SOC energy respectively. The drastic increase of $\frac{\tau_{so}^{-1}}{\tau_{tr}^{-1}}$ shown in Fig. 2f can be explained by the suppressed $E_g$ due to the gradually enhanced trigonal symmetry. Since $\mathcal{L}$ is larger than the Fermi energy (~ 0.2 eV from ref. [19]), the SOC has a non-negligible influence on the electron-lattice interaction process. The nearly 4 times enhancement of $\frac{\tau_{so}^{-1}}{\tau_{tr}^{-1}}$ further suggests the strengthened influence of SOC to the electron-lattice interaction. The spin-orbit scattering mixes spin and orbital states, allowing phonons to efficiently scatter electrons between states with different spin characters (spin-up to spin-down). This opens up new scattering channels that were previously forbidden or weak, effectively increasing the average EPC matrix element $<g^2>$. For example, the relatively high $T_c$ of 7.2 K in Pb is attributed to the strong SOC and its enhancement of EPC [49]. Therefore, we conjecture that the enhanced spin-orbit scattering in the (111) KTO systems increases the $T_c$ by providing a large $<g^2>$. Additionally, it is also possible that the trigonal symmetry together with strong SOC induce Fermi surface topology change, such as the emergence of flat band, to enhance the SC. However, previous density functional theory (DFT) calculations [47,48] did not report flat bands in the (111) KTO surface or heterostructure. Thus, we leave the possibility of SC enhancement driven by trigonal band structure for future investigations.

The presence of unconventional superconductivity in this system is intriguing, for which we propose the following explanation. As shown in Fig. 4b, due to the competition of the two types of symmetry, i.e. $O_h$ and $3m$, the Pomeranchuk-type Fermi surface instability could be trigged [50] which induces a quantum critical point (QCP) in the phase diagram. Indeed, the non-Fermi liquid behavior is observed (Fig. 1b) in all samples, demonstrating the existence of QCP. Because of the QCP, the long-range correlation (such as coulomb interaction) exists in a large finite temperature range. Furthermore, in samples from #6 to #1 with increasing $T_c$, the $N$ values in the $R \propto T^N$ relation evolves from ~ 1.53 to ~ 1.15. Although $N \sim 1.5$ may be attributed to quantum fluctuations with electron-electron interactions [51], the lower value of $N \sim 1.15$ in samples #1 and #2—which nearly exhibits a linear $R \propto T$ dependence—cannot be explained by this theory. The linear $R \propto T$ behavior is well-known in the normal state of high-$T_c$ cuprates, which has been extensively studied [52,53]. Recent theoretical calculations have revealed that the electron-lattice umklapp scattering via a critical bosonic mode (induced by the QCP) can lead to the linear $R \propto T$ relation [50], implying the approach to QCP in sample #1 and #2. Therefore, the existence of QCP is consistent with the measured normal state transport behavior such that the SC in the enhanced $T_c$ interface can be categorized as unconventional.

## Conclusion

In summary, we demonstrate that the synergetic interplay of trigonal crystal field and strong SOC account for enhancement of SC at the LAO/KTO (111) interface. The trigonal symmetry enhancement is observed from both the superconducting thickness reduction and the interface-sensitive nonlinear SFG spectroscopy. The enhanced SOC influence on the electron-lattice interaction is evidenced from the significantly increased spin-orbit scattering rate in samples of enhanced SC. Furthermore, the normal state resistance measurements reveal comprehensive evidences of both unconventional superconductivity and existence of QCP in the system. Therefore, our results illustrate the physical mechanism of the enhanced SC in KTO (111) interface, which also provides a plausible explanation for the distinct orientation-dependent SC in KTO interface families.

# Methods

**Sample fabrications** Prior to film growth, Hall bar patterns were defined on KTO (111) single-crystal substrates (5 × 5 × 0.5 mm³) using standard UV lithography. Subsequently, an approximately 40-nm-thick, oxygen-rich LAO film was deposited by pulsed laser deposition at room temperature under an oxygen partial pressure of 5 Pa to serve as a hard mask. The deposition employed a 248-nm KrF excimer laser with a fluence of ~1 J cm$^{-2}$ and a repetition rate of 2 Hz. The subsequent removal of this hard mask defined the active Hall-bar region on the uncovered KTO surface. Next, a 10-nm-thick, oxygen-deficient LAO film was deposited under an oxygen partial pressure of 5×10$^{-5}$ Pa and at substrate temperatures ranging from room temperature to 600 °C, using the same laser conditions. This process formed the Hall-bar-patterned 2DEGs at the LAO/KTO interfaces. Finally, to ensure Ohmic contacts, Pt electrodes were fabricated at the interfaces using overlay lithography and ion milling.

**STEM measurements** Cross-sectional specimens for scanning transmission electron microscopy (STEM) were prepared with Focused Ion Beam (FIB) (FEI Helios 600i) using lift-out method. The HAADF-STEM images were performed on a double aberration corrected field-emission STEM (JEOL Grand ARM 300F) operated at 300 kV. The spot size of 8C was used, and the convergence semi-angle of 22 mrad. HAADF imaging was conducted using the detection semi-angle range of 54-220 mrad. To improve the signal-to-noise ratio and minimize image drift, 5 frames (1024 × 1024 pixels, pixel dwell time of 4 μs) were collected and averaged. EDS spectral map-ping was conducted at a frame size of 256 × 256 pixels and a pixel dwell time of 10 μs. Spectral image collection took about 300 s to reach the targeted signal-to-noise ratio. All EDS images were processed via Wiener filtering.

**Magneto-electrical transport measurements** The Hall-bar-patterned 2DEGs were wire-bonded to the measurement setup using aluminum wires via ultrasonic bonding. Electrical transport measurements were carried out in two systems: a Physical Properties Measurement System (PPMS, Quantum Design) for the temperature range of 1.8 K to 300 K, and a commercial cryostat with a ³He insert (Oxford Instruments Teslatron PT) for temperatures down to 300 mK. A standard low-frequency lock-in technique was employed with an excitation current of 0.5 μA, particularly for the ultralow-temperature measurements.

The parameters $d_c$ and $d_{sc}$ are derived from the single-band Drude model and the G-L model, respectively. As Fig. 4b shows, two nearly-degenerate bands lie close to the Fermi level. A straightforward correction to the single band model is, therefore, to double the density of states on the 2D Fermi surface. This modifies the Fermi wave vector from $k_F = \sqrt{2\pi n_{2D}}$ to $k_F = \sqrt{\pi n_{2D}}$. According to Equation (1), $d_c$ is proportional to the mean free path, $l$. Given that $l \propto v_F = \hbar k_F/m^*$, a correction of $\frac{1}{\sqrt{2}}$ must be applied for $d_c$. Interestingly, the corrected $d_c$ (e.g. 7.6 nm in sample #6) is close to the correponding $d_{sc}$ (6.4 nm). The consistency between $d_c$ and $d_{sc}$, obtained from two distinct models, strongly validates the reliability of our layer thickness analyses.

The spin-orbit scattering time, $\tau_{so}$, was extracted by fitting the $\mu_0 H_{c2\parallel}(T)$ data to the KLB model [33,34], which is expressed by the following equation:

$$ln\frac{T}{T_c} = \psi\left(\frac{1}{2}\right) - \psi\left(\frac{1}{2} + \frac{3\tau_{so}\mu_B^2 H_{c2\parallel}^2}{4\pi\hbar k_B T}\right) \quad (5)$$

where $\psi(x)$ is the digamma function. As shown in Fig. E9, the KLB model provides a nice fit to the $\mu_0 H_{c2\parallel}(T)$ curves. Furthermore, we employed the modified WHH model [41,42] to determine the orbital depairing parameter $\alpha$ and the spin-orbit scattering parameter $\lambda_{so}$, defined as $\alpha = \frac{\hbar}{2m^*D}$ and $\lambda_{so} = \frac{2\hbar}{3\pi k_B T_c \tau_{so}}$, where $D$ is the diffusion coefficient. They can be derived from the following equations:

$$\left.\frac{-d(\mu_0 H_{c2\perp})}{dT}\right|_{T=T_c} = \frac{4k_B}{\pi D e} \quad (6)$$

$$\mu_0 H_{c2\parallel}(T=0) = \sqrt{\frac{1.76\hbar k_B T_c}{3\mu_B^2 \tau_{so} + \frac{D(d_{sc}e)^2}{3}}} \quad (7)$$

The values of $\alpha$, $\lambda_{so}$, $\tau_{tr}$, and $\tau_{so}$ for all six samples are summarized in Fig. E10.

**SFG measurements** The SFG experiments were carried out with a femtosecond broadband laser system. Briefly, a Ti:Sapphire amplifier (Spitfire ACE, Spectra Physics) produces ~7 W of 800 nm, 35 fs pulses at 2 kHz repetition rate. 40% of the 800 nm beam passed through a beam-splitter to pump an optical parameter amplifier followed by a difference frequency generation stage (TOPAS-C, Spectra Physics). The rest was reflected from a Bragg filter (N013-14-A2, OptiGrate) to generate a narrow band beam of ~0.5 nm bandwidth. The broadband IR and narrowband 800 nm pulses overlapped at the sample surface with incident angles of 57° and 45°, respectively. The SF signal was collected along the reflected direction by a spectrograph (Acton SP2300) and a CCD camera (Princeton Instruments PyLoN 1340 × 100).

## Acknowledgements

We acknowledge insightful discussions with Prof. Yanwu Xie and Prof. Yi Zhou. The work was financially supported by National Key R&D Program of China (Nos. 2022YFA1403000, 2023YFA1406301), the National Science Foundation of China (Nos. 12250710675, 12250002, 12074276), the Science and Technology Commission of Shanghai Municipality (Nos. 23JC1400400). The research used resources from Analytical Instrumentation Center (#SPST-AIC10112914) and Soft Matter Nanofab (SMN180827) in ShanghaiTech University.

## Author contributions

J.L., L.C., and X.Z. fabricated samples, performed transport measurements, and conducted the associated data analysis, with support from M.Z., F.Y., and Q.W.. T.L., J.L., S.Z., and W.L. conducted the SFG measurements and analyzed the corresponding data. P.C. and J.Z. carried out the STEM-EDS measurement and analysis. The theoretical analysis was provided by J.Z., X.Z., and J.K. All authors discussed the results and their interpretation.

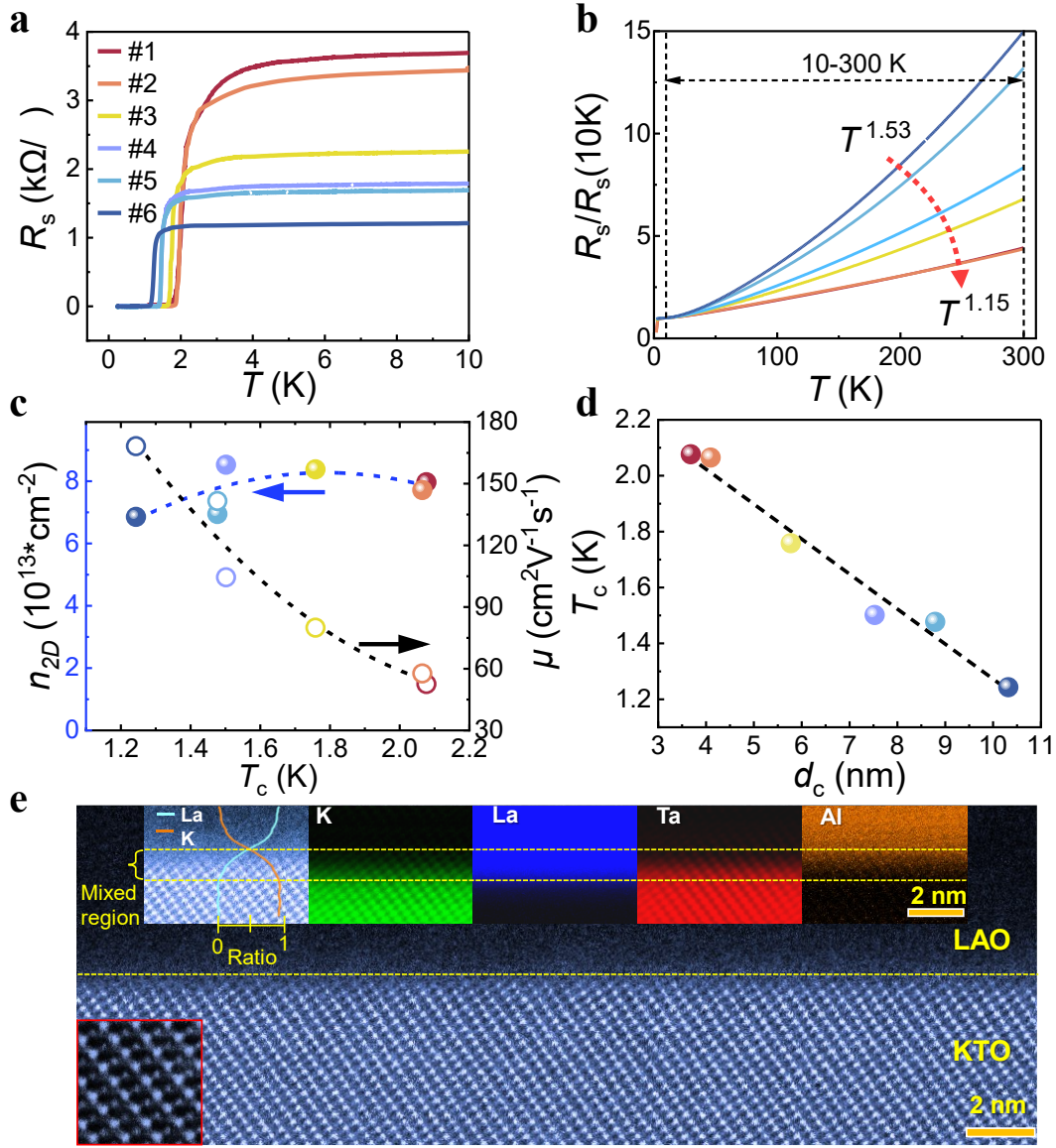

**Figure 1: Electrical transport and structural characterization of the LAO/KTO (111) samples. a** Sheet resistance ($R_s$) as a function of temperature for LAO/KTO interfaces measured below 10 K. **b** Sheet resistance normalized by its value at 10K. A power law fit of $R_s \propto T^N$ in the range of 10-300 K reveals the exponent $N$ decreasing from 1.53 to 1.15 as $T_c$ increases. **c** 2D carrier density and mobility measured at 12 K. **d** The relationship between $T_c$ and the conductive layer thickness ($d_c$). **e** The HADDF-STEM image of sample #1 viewed along the [11-2] zone axis, alongside the corresponding EDS elemental mapping of the LAO/KTO interface. The yellow dash lines mark the intermixed region where La substitutes K sites on the KTO side of the interface.

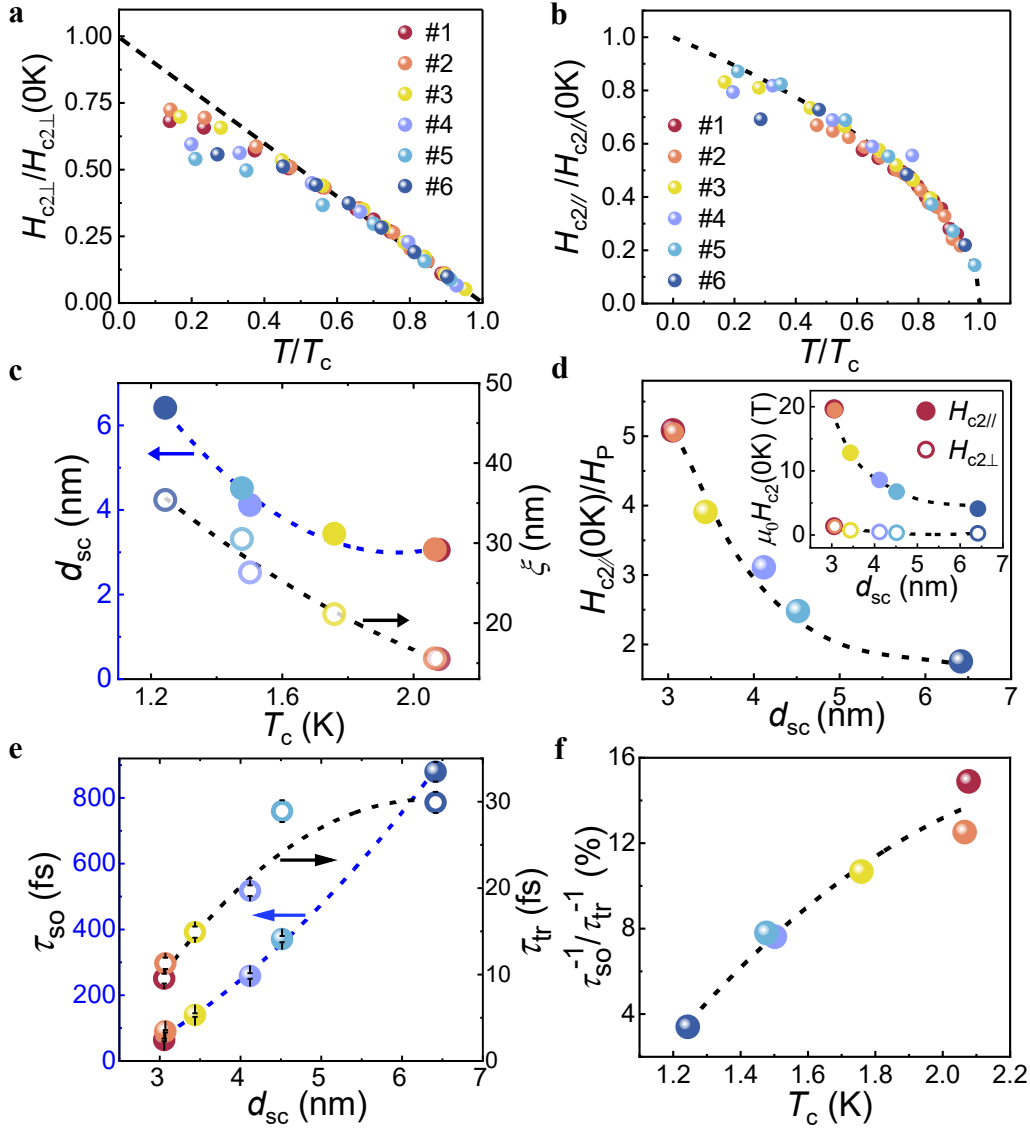

**Figure 2: Magneto-transport characterization. a-b** Normalized upper critical fields as a function of temperature under **a** perpendicular and **b** parallel fields. The dashed lines are fits based on the 2D G-L model, from which the zero-temperature $H_{c2//}(0K)$ and $H_{c2\perp}(0K)$ are derived. **c** The coherence length and superconducting layer thickness derived from the G-L model. **d** The ratio of $H_{c2//}(0K)/H_P$ as a function of $d_{sc}$. The inset shows the $d_{sc}$-dependent $H_{c2//}(0K)$ and $H_{c2\perp}(0K)$ derived from the G-L fits. **e** The transport scattering time ($\tau_{tr}$) and spin-orbit scattering time ($\tau_{so}$) (obtained from the KLB model) as functions of $d_{sc}$. The error bars are represented by short bars within the data points. **f** The relationship between the ratio $\frac{\tau_{so}^{-1}}{\tau_{tr}^{-1}}$ and $T_c$.

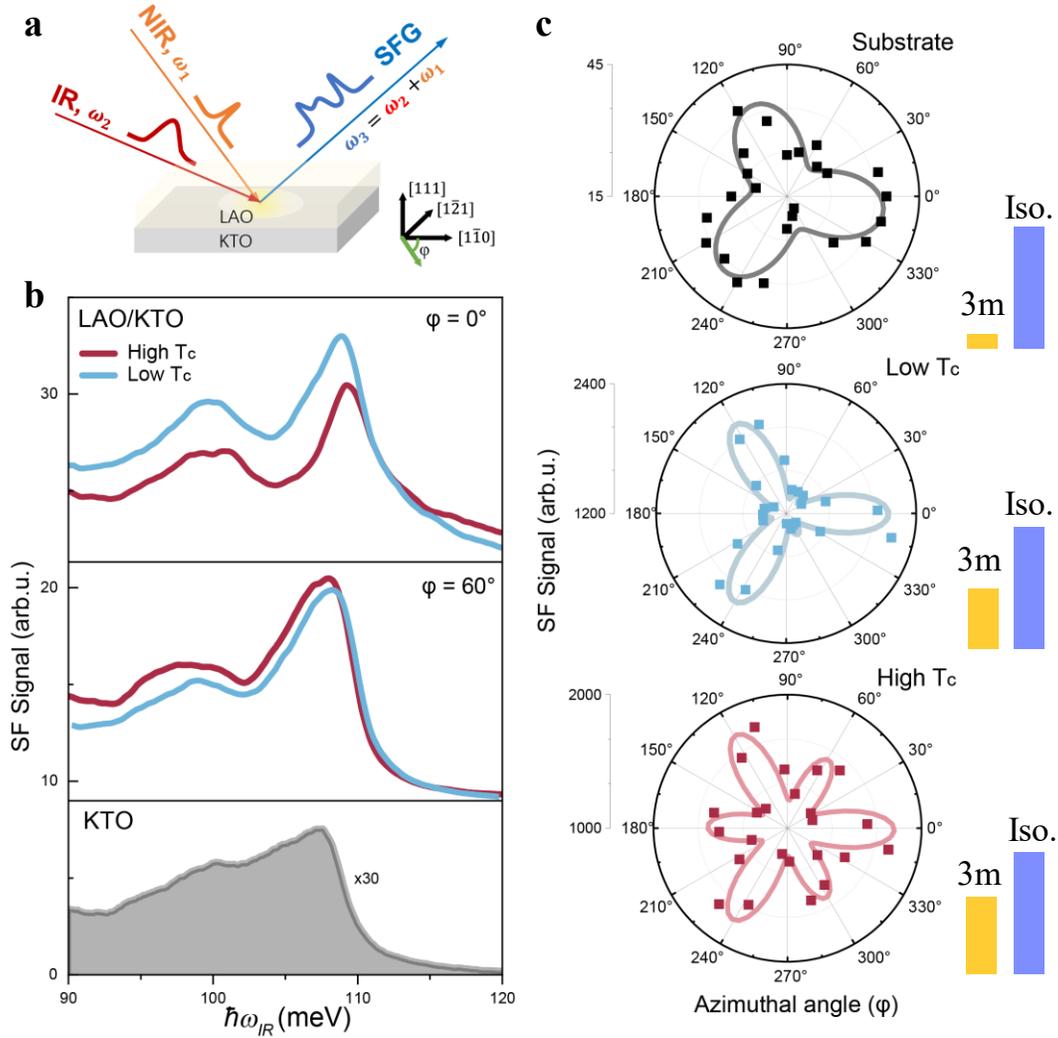

**Figure 3: Sum-frequency generation (SFG) vibrational spectroscopy of the KTO (111) surface and a-LAO/KTO interfaces. a** Experimental optical path diagram: Spatially and temporally overlapping infrared (IR) and near-infrared (NIR) beams from a femtosecond pulsed laser system generate the sum-frequency signal at the sample. **b** SFG spectra of bare KTO (111) surface (gray fill) and LAO/KTO interfaces with different $T_c$ (colored curves, taken at $\varphi = 0°$ and $60°$). The peak at 107 meV corresponds to the $LO_4$-derived surface phonon mode, while the low-frequency band at 100 meV likely originates from oxygen vacancy-related local modes. **c** Integrated SFG intensity (90-120 meV) as a function of azimuthal angle $\varphi$ for three samples, all showing threefold symmetry characteristic of the (111) crystal plane. The right-hand side of each pole figure shows the trigonal contribution (term A in Equation (4)) compared to the isotropic contribution (term B in Equation (4)).

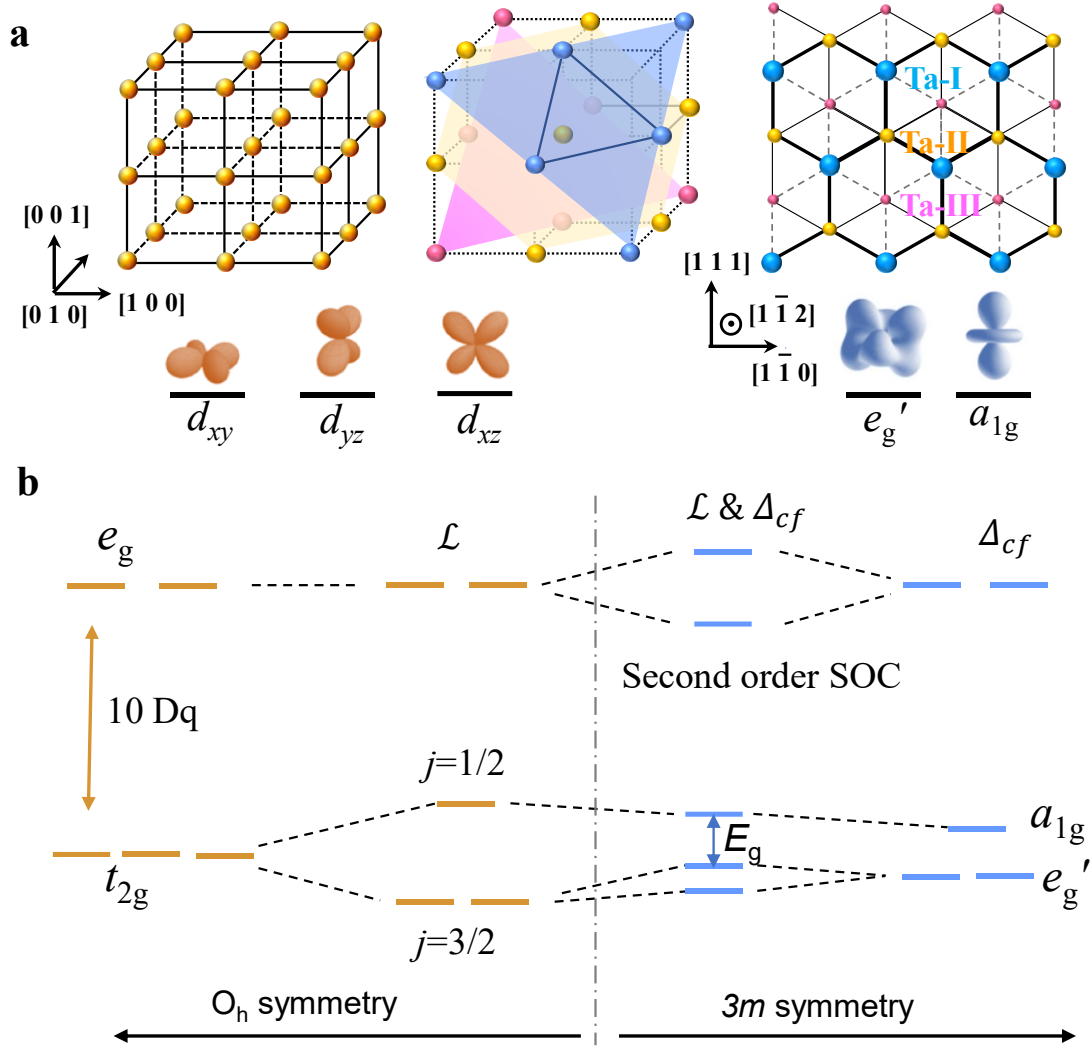

**Figure 4: a** Schematic illustrations of the bulk cubic perovskite structure of KTO (left), a slab sliced along the (111) direction (middle), and a top-down view of the (111) slab (right). The bottom panels schematically show the $t_{2g}$ orbitals in the $O_h$ symmetry (orange lobes) and the trigonal symmetry (3m) (blue lobes). **B** Hierarchical splitting of the Ta 5d orbitals. In a cubic structure with $O_h$-symmetric crystal field (10Dq), the 5d orbitals split into $e_g$ and $t_{2g}$ manifolds. When SOC ($\mathcal{L}$) is included, the $t_{2g}$ manifold further splits into two levels with effective total angular momentum $j=1/2$ and 3/2. Under a trigonal crystal field ($\Delta_{cf}$), the $t_{2g}$ manifold splits into $a_{1g}$ and $e'_g$ levels. With both SOC and the 3m symmetry are present, all degeneracies are lifted: the $t_{2g}$ manifold splits into three levels and the $e_g$ manifold splits into two.